\documentclass[aps,twocolumn,floats,pre,floatfix]{revtex4}
\usepackage{graphicx,graphicx,epsfig}
\usepackage{amssymb}
\usepackage{epsf,wrapfig} 	

\begin{document}

\title{Neural coding of a natural stimulus ensemble:\\
Uncovering information at sub--millisecond resolution}

\author{Ilya Nemenman,$^a$ Geoffrey D. Lewen,$^b$ William Bialek$^c$ and Rob R. de Ruyter van Steveninck$^d$}

\affiliation{$^a$Computer,  Computational and Statistical Sciences Division\\
  Los Alamos National Laboratory, Los Alamos, New Mexico 87545\\
  $^b$The Hun School of Princeton, 176 Edgerstoune Road, Princeton, New Jersey 08540\\
  $^c$Joseph Henry Laboratories of Physics and Lewis--Sigler Institute for Integrative Genomics\\
  Princeton University, Princeton, New Jersey 08544\\
  $^d$Department of Physics, Indiana University, Bloomington, Indiana
  47405}

\begin{abstract}
  Our knowledge of the sensory world is encoded by neurons in
  sequences of discrete, identical pulses termed action potentials or
  spikes.  There is persistent controversy about the extent to which
  the precise timing of these spikes is relevant to the function of
  the brain.  We revisit this issue, using the motion--sensitive
  neurons of the fly visual system as a test case.  New experimental
  methods allow us to deliver more nearly natural visual stimuli,
  comparable to those which flies encounter in free, acrobatic flight,
  and new mathematical methods allow us to draw more reliable
  conclusions about the information content of neural responses even
  when the set of possible responses is very large.  We find that
  significant amounts of visual information are represented by details
  of the spike train at millisecond and sub--millisecond precision,
  even though the sensory input has a correlation time of $\sim
  60\,{\rm ms}$; different patterns of spike timing represent distinct
  motion trajectories, and the absolute timing of spikes points to
  particular features of these trajectories with high precision.
  Under these naturalistic conditions, the system continues to
  transmit more information at higher photon flux, even though
  individual photoreceptors are counting more than one million photons
  per second, and removes redundancy in the stimulus to generate a
  more efficient neural code.
\end{abstract}

\date{\today}

\maketitle

\section{Introduction}

Throughout the brain, information is represented by discrete
electrical pulses termed action potentials or `spikes' \cite{spikes}.
For decades there has been controversy about the extent to which the
precise timing of these spikes is significant: should we think of each
spike arrival time as having meaning down to millisecond precision
\cite{mackay+mcculloch_52,abeles_82,strong+al_98}, or does the brain
only keep track of the number of spikes occurring in much larger
windows of time?  Is precise timing relevant only in response to
rapidly varying sensory stimuli, as in the auditory system
\cite{carr_93}, or can the brain construct specific patterns of spikes
with a time resolution much smaller than the time scales of the
sensory and motor signals that these patterns represent
\cite{abeles_82,hopfield_95}?  Here we address these issues using the
motion--sensitive neurons of the fly visual system as a model
\cite{hausen_84}.

We bring together new experimental methods for delivering truly
naturalistic visual inputs \cite{lewen+al_01} and new mathematical
methods that allow us to draw more reliable inferences about the
information content of spike trains
\cite{nemenman+al_02,nemenman_02,nemenman+al_04}.  We find that as we
improve our time resolution for the analysis of spike trains from
$2\,{\rm ms}$ down to $0.2\,{\rm ms}$ we reveal nearly one--third more
information about the trajectory of visual motion.  The natural
stimuli used in our experiments have essentially no power above
$30\,{\rm Hz}$, so that the precision of spike timing is not a
necessary correlate of the stimulus bandwidth; instead the different
patterns of precise spike timing represent subtly different
trajectories chosen out of the stimulus ensemble.  Further, despite
the long correlation times of the sensory stimulus, segments of the
neural response separated by $\sim 30\,{\rm ms}$ provide essentially
independent information, suggesting that the neural code in this
system achieves decorrelation \cite{barlow_59,barlow_61} in the time
domain.  This decorrelation is not evident in the time dependent spike
rate alone, but the time scale for the independence of information
does match the time scale on which visual motion signals are used to
guide behavior \cite{land+collett_74}.

\section{Posing the problem}

\begin{figure*}[bht]
\vskip -0.25 in
\includegraphics[width = 0.48\linewidth,angle=270]{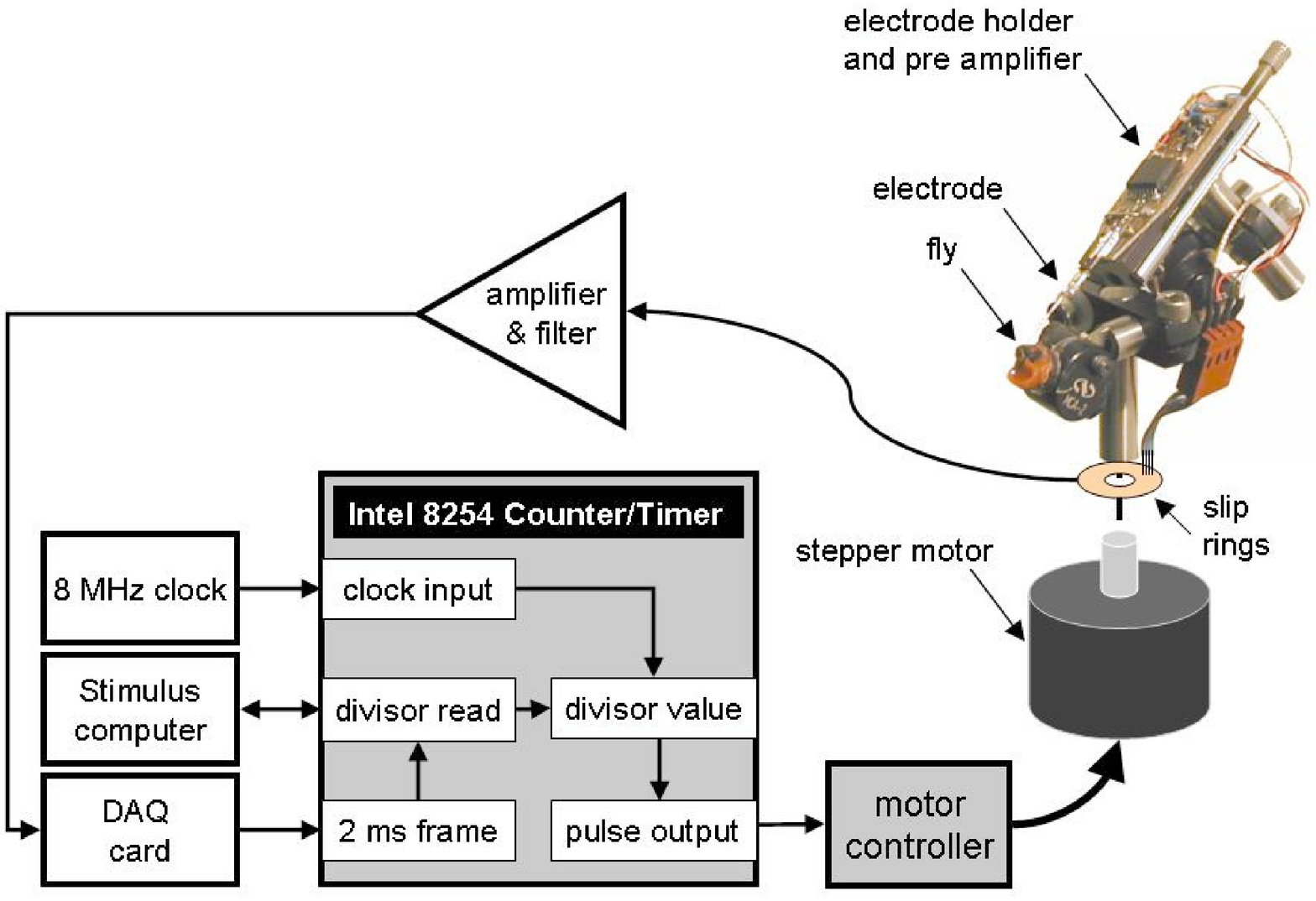}\hfill\includegraphics[width = 0.48\linewidth]{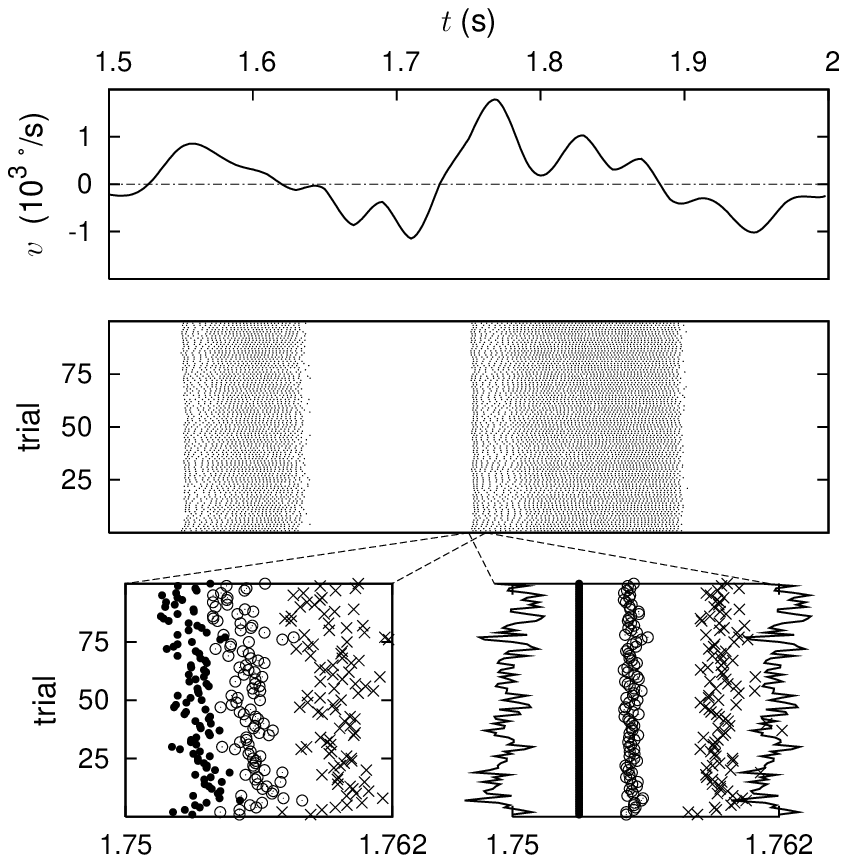}
\caption{Neural responses to a natural stimulus ensemble.  At left we
  show a schematic of the experimental setup (see Methods for
  details).  A fly is immobilized with wax, its body in a plastic
  tube, with its head protruding. Through a small hole in the back of
  the head an electrode is inserted to record extracellular potentials
  from H1, a wide field neuron sensitive to horizontal motion. This
  signal is amplified, fed through a slip ring system to a second
  stage amplifier and filter, and recorded by a data acquisition card.
  In synchrony with its master timer clock, the DAQ card generates a
  $500\,{\rm Hz}$ frame clock signal. Every $2\,{\rm ms}$, through a
  bidirectional parallel port, this clock triggers a successive read
  of a divisor value from a file stored in the stimulus laptop
  computer. The Intel 8254 Counter/Timer chip uses this divisor value
  to divide down the pulse frequency of a free running $8\,{\rm MHz}$
  clock. In this way, in each successive $2\,{\rm ms}$ interval, and
  in strict synchrony with the data taking clock, a defined and evenly
  spaced burst of pulses is produced. These pulses drive the stepper
  motor, generating the angular velocity signal.  A brief segment of
  this motion stimulus is shown in the top right panel, below which we
  plot a raster of action potentials from H1 in response to 100
  repetitions of this stimulus.  At bottom we expand the scale to
  illustrate (at left) that individual spikes following a transition
  from negative to positive velocity jitter from trial to trial by
  $\sim 1\,{\rm ms}$: the standard deviations of spike times shown
  here are $0.72\,{\rm ms}$ for the first spike ($\bf\cdot$),
  $0.81\,{\rm ms}$ for the second spike ($\circ$), and $1.22\,{\rm
    ms}$ for the third spike ($\times$).  When we align the first
  spikes in this window, we see (at right) that the jitter of
  interspike intervals is even smaller, $0.21\,{\rm ms}$ for the first
  interval and $0.69\,{\rm ms}$ for the second interval.  Our
  challenge is to quantify the information content of such precise
  responses.}
\label{big_expt}
\end{figure*}

Flies exhibit a wide variety of visually guided behaviors, of which
perhaps the best known is the optomotor response, in which visual
motion drives a compensating torque, stabilizing straight flight
\cite{reichardt+poggio_76}.  This system offers many advantages for
the exploration of neural coding and computation: there is a small
groups of identified, wide--field motion--sensitive neurons
\cite{hausen_84} that provide an obligatory link in the process
\cite{hausen+wehrhahn_83}, and it is possible to make very long,
stable recordings from these neurons as well as to characterize in
detail the signal and noise properties of the photoreceptors that
provide the input data for the computation.  In free flight, the
trajectory of visual motion is determined largely by the fly's own
motion through the world, and there is a large body of data on flight
behavior under natural conditions
\cite{land+collett_74,wagner_86,schlistra+hateren_99,hateren+schlistra_99},
offering us the opportunity to generate stimuli that approximate those
experienced in nature.  But the natural visual world of flies involves
not only the enormous angular velocities associated with acrobatic
flight; natural light intensities and the dynamic range of their
variations also are very large, and the fly's compound eyes are
stimulated over more than $2\pi$ steradians; all of these features are
difficult to replicate in the laboratory \cite{ruyter+al_01}.  As an
alternative, we have moved our experiments outside \cite{lewen+al_01},
so that flies experience the scenes from the region in which they were
caught.  We record from a single motion--sensitive cell, H1, while we
rotate the fly along trajectories that are modeled on the natural
flight trajectories (see Methods for details).  For other approaches
to the delivery of naturalistic stimuli in this system see
\cite{hateren+al_05}.

A schematic of our experiment, and an example of the data we obtain,
are shown in Fig \ref{big_expt}.  We see qualitatively that the
responses to natural stimuli are very reproducible, and we can point
to specific features of the stimulus---such as reversals of motion
direction---that generate individual spikes and interspike intervals
with better than millisecond precision.  The challenge is to quantify
these observations: do precise and reproducible patterns of spikes
occur just at some isolated moments, or does looking at the spike
train with higher time resolution generally provide more information
about the visual input?

Precise spike timing endows each neuron with a huge ``vocabulary'' of
responses \cite{spikes, mackay+mcculloch_52}, but this potential
advantage in coding capacity creates challenges for experimental
investigation. If we look with a time resolution of $\tau = 1\,{\rm
  ms}$, then in each bin of size $\tau$ we can see either zero or one
spike; across the behaviorally relevant time scale of $30\,{\rm ms}$
the neural response thus can be described as a 30--bit binary word,
and there are $2^{30}$, or roughly one billion such words.  Although
some of these responses never occur (because of refractoriness) and
others are expected to occur only with low probability, it is clear
that if precise timing is important then neurons can generate many
more meaningfully distinguishable responses than the number that we
can sample in realistic experiments.

\begin{figure}
\centerline{\includegraphics[width = 3.5 in]{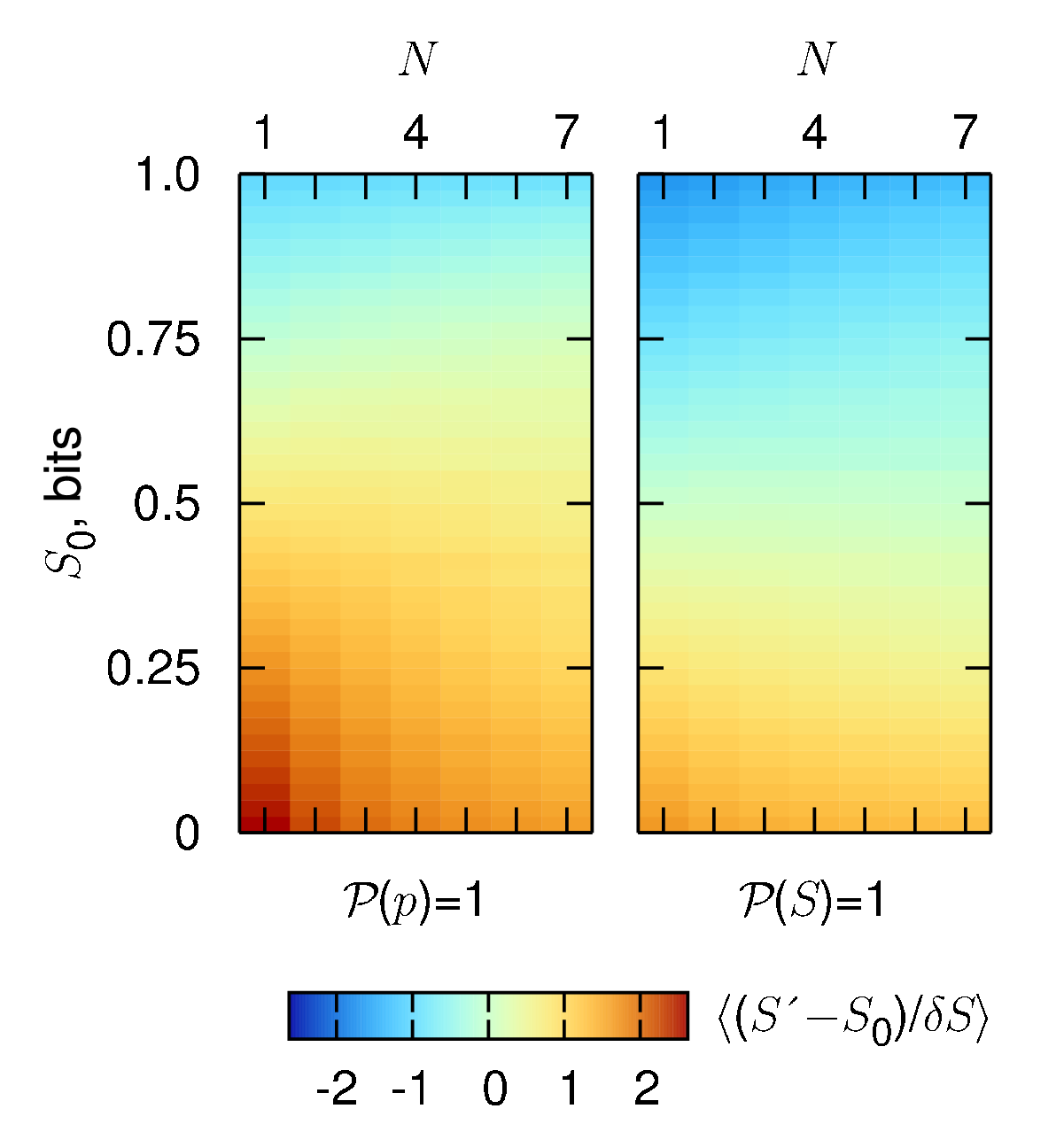}}
\caption{Systematic errors in entropy estimation. We consider a coin
  with unknown probability $p$ of coming up heads; from $N$ coin flips
  we try to estimate the entropy $S = -p\log_2 p - (1-p)\log_2(1-p)$;
  see Methods for details of the calculations.  At left, we make
  Bayesian estimates starting from the prior hypothesis that all
  values of $p$ are equally likely, ${\cal P}(p) = 1$.  We show how
  the best estimate $S'$ differs from the true value $S_0$ when this
  deviation is measured in units of the estimated error bar $\delta S$
  (posterior standard deviation).  For small numbers of samples, the
  best estimate is systematically in error by more than two times the
  size of the error bar, so we would have false confidence in a wrong
  answer, even at intermediate values of the entropy which are most
  relevant for real data.  At right, we repeat the same procedure but
  with a prior hypothesis that all possible value of the entropy are
  equally likely, ${\cal P}(S) = 1$.  Systematic errors still appear,
  but they are more nearly compatible with the error bars, even at
  small $N$, and especially in the range of entropies which is
  relevant to our experiments.}
\label{errors}
\end{figure}

Can we make progress on assessing the content and meaning of neural
responses even when we can't sample all of them? Some hope is provided
by the classical problem of how many people need to be present in a
room before there is a reasonable chance that they share a birthday.
This number, $N\sim 23$, is vastly less than the number of possible
birthdays, $K = 365$.  Turning this argument around, if we didn't know
the number of possible birthdays we could estimate it by polling $N$
people and checking the frequency of coincidences.  Once $N$ is large
enough to generate several coincidences we can get a pretty good
estimate of $K$, and this happens when $N \sim \sqrt{K} \ll K$. Some
years ago Ma proposed that this coincidence counting method be used to
estimate the entropy of physical systems from molecular dynamics or
Monte Carlo simulations \cite{ma_81} (see also Ref \cite{serber_73}).
If these arguments could be generalized, it would become feasible to
estimate the entropy and information content of neural responses even
when experiments provide only a sparse sampling of these responses.
The results of Ref \cite{nemenman+al_02,nemenman_02} provide such a
generalization.

To understand how the methods of Ref \cite{nemenman+al_02} generate
more accurate entropy estimates from small samples, it is useful to
think about the simpler problem of flipping a coin under conditions
where we don't know the probability $p$ that it will come up heads.
One strategy is to count the number of heads $n_H$ that we see after
$N$ flips, and identify $p=n_H/N$; if we then use this ``frequentist''
or maximum likelihood estimate to compute the entropy of the
underlying distribution, it is well known that we will underestimate
the entropy systematically
\cite{miller_55,treves+panzeri_95,paninski_03}.  Alternatively, we
could take a Bayesian approach and say that a priori all values of $0
< p <1$ are equally likely; the standard methods of Bayesian
estimation then will generate a mean and an error bar for our estimate
of the entropy given $N$ observations.  As shown in Fig \ref{errors},
this procedure actually leads to a systematic {\em overestimate} of
the entropy in cases where the real entropy is not near its maximal
value.  More seriously, this systematic error is larger than the error
bars that emerge from the Bayesian analysis, so we would be falsely
confident in the wrong answer.

Figure \ref{errors} also shows us that if we use a Bayesian approach
with the a priori hypothesis that all values of the entropy are
equally likely, then (and as far as we know, only then) we find
estimates such that the systematic errors are comparable to or smaller
than the error bars, even when we have seen only one sample.  Thus the
problem of systematic errors in entropy estimation is not, as one
might have thought, the problem of not having seen all the
possibilities; the problem rather is that seemingly natural and
unbiased prior hypotheses about the nature of the underlying
probabilities correspond to highly biased hypotheses about the entropy
itself, and this problem gets much worse when we consider
distributions over many alternatives.  The strategy of Ref
\cite{nemenman+al_02} thus is to construct, at least approximately, a
`flat prior' on the entropy (see Methods for details).  The results of
Ref \cite{nemenman+al_04} demonstrate that this procedure actually
works for both simulated and real spike trains, where `works' means
that we generate estimates that agree with the true entropy within
error bars even when the number of samples is much smaller than the
number of possible responses.  As expected from the discussion of the
birthday problem, what is required for reliable estimation is that the
number of coincidences be significantly larger than one
\cite{nemenman_02}.

\begin{figure}
\includegraphics[width = 0.95\linewidth]{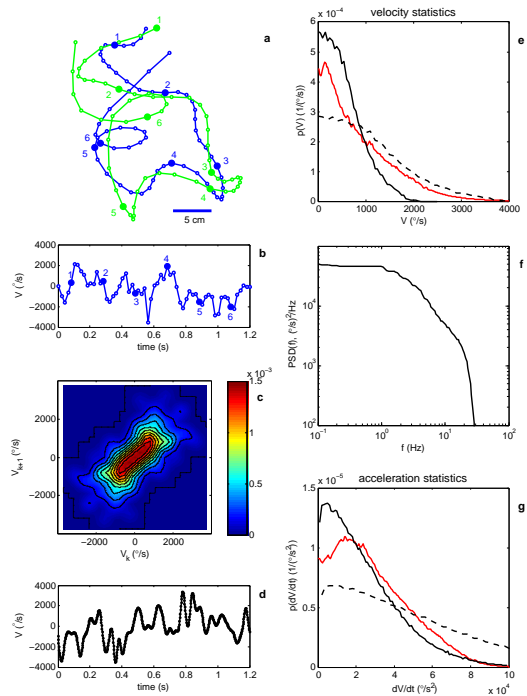}
\vskip -0.25 in
\caption{Constructing a naturalistic stimulus. (a) Digitized version
  of original video tracking data by Land and Collett
  \cite{land+collett_74}. The panel shows traces of a leading fly
  (blue) and a chasing fly (green). Successive points along the
  trajectories are recorded at $20\, {\rm ms}$ intervals. Every tenth
  point along each trajectory is indicated by a number. From these
  traces we estimate rotational velocities of the body axis by
  calculating the angular change in orientation of the trajectory from
  one point in the sequence to the next, and dividing by $20\, {\rm
    ms}$. The result of this calculation for the leading fly is shown
  in panel (b).  (c) From these data (on both flies) we construct a
  joint distribution, $P(V_k,V_{k+1})$, of successive velocities taken
  $20\, {\rm ms}$ apart.  (d) Short sample of a trajectory constructed
  using the distribution in (c) as a Markov process, and then
  interpolating the velocity trace to $2\,{\rm ms}$ resolution.  (e)
  Probability densities of angular velocity generated from this Markov
  process (black dashed line) and scaled down by a factor of two
  (black line) to avoid destabilizing the experiment; distributions
  are symmetric and we show only positive velocities. For comparison
  we show (red line) the distribution of angular velocities recorded
  for head motion of {\em Calliphora} during episodes of saccadic
  turning \cite{hateren+schlistra_99}.  (f) Power spectrum of
  synthesized velocity signal, demonstrating the absence of power
  above $30\,{\rm Hz}$.  (g) As in (e) but for the accelerations.
  Note that the distribution of our synthesized and scaled signal is
  surprisingly close to that found for saccadic head motions.}
\label{stim}
\end{figure}

\section{Words, entropy and information}

\begin{figure*}
\includegraphics[width=0.8\linewidth]{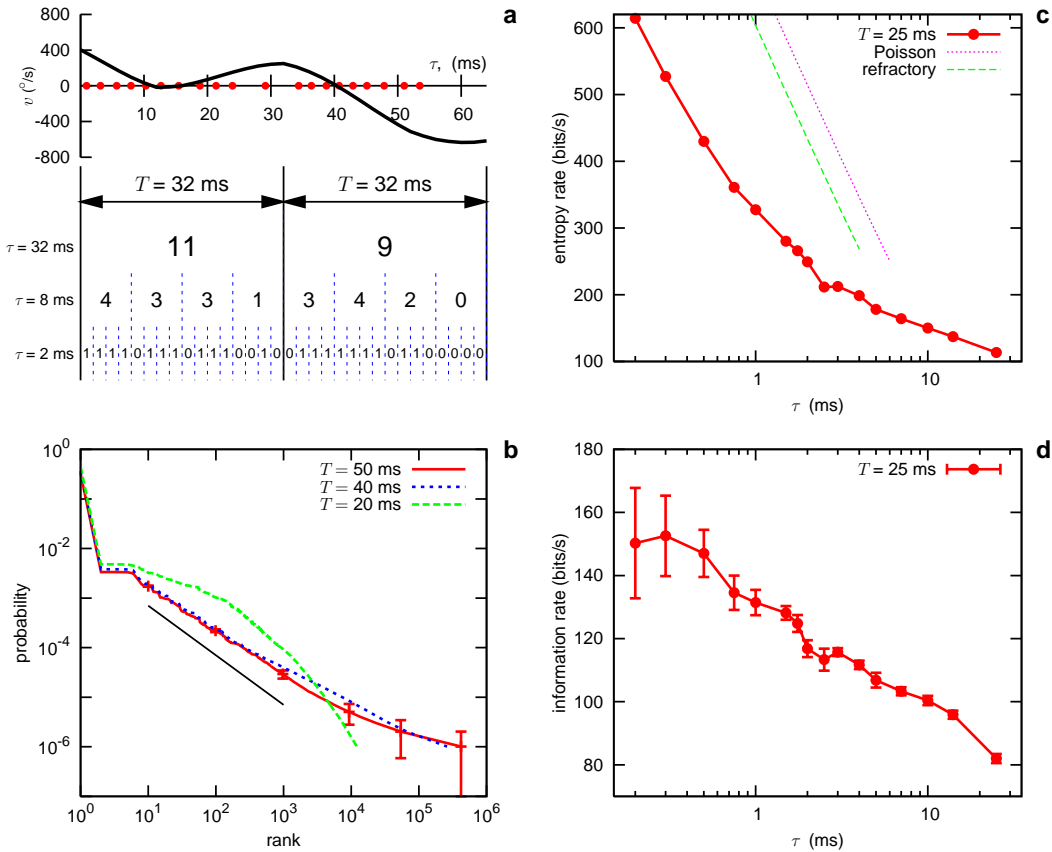}
\caption{Words, entropy and information in the neural response to
  natural signals.  (a) Schematic showing how we convert the sequence
  of action potentials into discrete `words'
  \cite{ruyter+al_97,strong+al_98}. At the top we show the stimulus
  and spike arrival times (red dots) in a $64\,{\rm ms}$ segment of
  the experiment.  We treat this as two successive segments of
  duration $T=32,{\rm ms}$, and divide these segments into bins of
  duration $\tau = 2$, $8$, or $32\,{\rm ms}$.  For sufficiently small
  $\tau$ (here, $\tau=2\,{\rm ms}$), each bin contains either zero or
  one spike, and so each neural response becomes a binary word with
  $T/\tau$ bits; larger values of $\tau$ generate larger alphabets,
  until at $\tau =T$ the response of the neuron is just the spike
  count in the window of duration $T$.  Note that the words are shown
  here as non--overlapping; this is just for graphical convenience.
  (b) The distribution of words with $\tau = 1\,{\rm ms}$, for various
  values of $T$; words are plotted in rank order.  We see that, for
  large $T$ ($T=40$ or $50\,{\rm ms}$) but not for small $T$
  ($T=20\,{\rm ms}$), the distribution of words has a large segment in
  which the probability of a word is $P\propto 1/{\rm rank}^\alpha$,
  corresponding to a straight line on this double logarithmic plot.
  Similar behavior is observed for words in English, with $\alpha =1$,
  which we show for comparison (solid line); this is sometimes
  referred to as Zipf's law \cite{zipf}.  (c) The entropy of a
  $T=25\,{\rm ms}$ segment of the spike train, as a function of the
  time resolution $\tau$ with which we record the spikes.  We plot
  this as an entropy rate, $S(T,\tau)/T$, in $\rm bits/s$; this value
  of $T$ is chosen because this is the time scale on which visual
  motion drives motor behavior \cite{land+collett_74}.  For comparison
  we show the theoretical results (valid at small $\tau$) for a
  Poisson process \cite{spikes}, and a Poisson process with a
  refractory period \cite{nemenman+al_04}, with spike rates and
  refractory periods matched to the data.  Note that the real spike
  train has significantly less entropy than do these simple models.
  In Ref \cite{nemenman+al_04} we showed that our estimation methods
  can recover the correct results for these models using data sets
  comparable in size to the one analyzed here; thus our conclusion
  that real entropies are smaller cannot be the result of
  undersampling.  Error bars are smaller than the data points.  (d)
  The information content of $T=25\,{\rm ms}$ words, as a function of
  time resolution $\tau$; again we plot this as a rate $R_{\rm
    info}(T,\tau ) = I(T,\tau)/T$, in $\rm bits/s$.}
\label{ent+inf_results}
\end{figure*}

Armed with tools that allow us to estimate the entropy of neural
responses, we first analyze a long experiment in which the fly
experiences a continuous trajectory of motion with statistics modeled
on those of natural flight trajectories (Fig \ref{stim}; see Methods
for details).  As shown in Fig \ref{ent+inf_results}a, we examine
segments of the response of duration $T$, and we break these segments
into discrete bins with time resolution $\tau$.  For sufficiently
small $\tau$ each bin either has one or zero spikes, and hence the
response becomes a binary word with $T/\tau$ bits, while in the
opposite limit we can let $\tau =T$ and then the response is the total
number of spikes in a window of size $T$; for intermediate values of
$\tau$ the responses are multi--letter words, but with larger than
binary alphabet when more than one spike can occur within a single
bin.  An interesting feature of these words is that they occur with a
probability distribution similar to the distribution of words in
English (Zipf's law; Fig \ref{ent+inf_results}b).  This Zipf--like
behavior emerges only for $T> 20\,{\rm ms}$, and was not observed in
experiments with less natural, noisy stimuli \cite{strong+al_98}.

With a fixed value of $T$, improving our time resolution (smaller
$\tau$) means that we distinguish more alternatives, increasing the
``vocabulary'' of the neuron.  Mathematically this means that the
entropy $S(T,\tau)$ of the neural responses is larger, corresponding
to a larger capacity for carrying information.  This is shown
quantitatively in Fig \ref{ent+inf_results}c, where we plot the
entropy rate, $S(T,\tau)/T$.  The question of whether precise spike
timing is important in the neural code is precisely the question of
whether this capacity is used by the system to carry information
\cite{mackay+mcculloch_52,strong+al_98}.

To estimate the information content of the neural responses, we follow
the strategy of Refs \cite{ruyter+al_97, strong+al_98}.  Roughly
speaking, the information content of the `words' generated by the
neuron is less than the total size of the neural vocabulary because
there is some randomness or noise in the association of words with
sensory stimuli.  To quantify this noise we choose a five second
segment of the stimulus, and then repeat this stimulus 100 times.  At
each moment $0<t<5\,{\rm s}$ in the cycle of the repeated stimulus, we
can look across the one hundred trials to sample the different
possible responses to the same input, and with the same mathematical
methods as before we use these samples to estimate the `noise entropy'
$S_n(T,\tau | t )$ in this `slice' of responses.  The information
which the responses carry about the stimulus then is given by
$I(T,\tau) = S(T,\tau ) - \langle S_n(T,\tau |T)\rangle_t$, where
$\langle \cdots \rangle_t$ denotes an average over time $t$, which
implicitly is an average over stimuli.  It is convenient to express
this as an information rate $R_{\rm info}(T,\tau ) = I(T,\tau)/T$, and
this is what we show in Fig \ref{ent+inf_results}d, with $T=25\,{\rm
  ms}$ chosen to reflect the time scale of behavioral decisions
\cite{land+collett_74}.

The striking feature of Fig \ref{ent+inf_results}d is the growth of
information rate with time resolution.  We emphasize that this
measurement is made under conditions comparable to those which the fly
encounters in nature---outdoors, in natural light, moving along
trajectories with statistics similar to those observed in free flight.
Thus, under these conditions, we can conclude that the fly's visual
system carries information about motion in the timing of spikes down
to sub--millisecond resolution.  Quantitatively, information rates
double as we increase our time resolution from $\tau = 25\,{\rm ms}$
to $\tau = 0.2\,{\rm ms}$, and the final $\sim 30\%$ of this increase
occurs between $\tau = 2\,{\rm ms}$ and $\tau = 0.2\,{\rm ms}$.  In
the behaviorally relevant time windows \cite{land+collett_74}, this
$30\%$ extra information corresponds to a almost a full bit from this
one cell, which would provide the fly with the ability to distinguish
reliably among twice as many different motion trajectories.

\section{What do the words mean?}

\begin{figure*}[bht]
\centerline{\includegraphics[width = 0.4\linewidth]{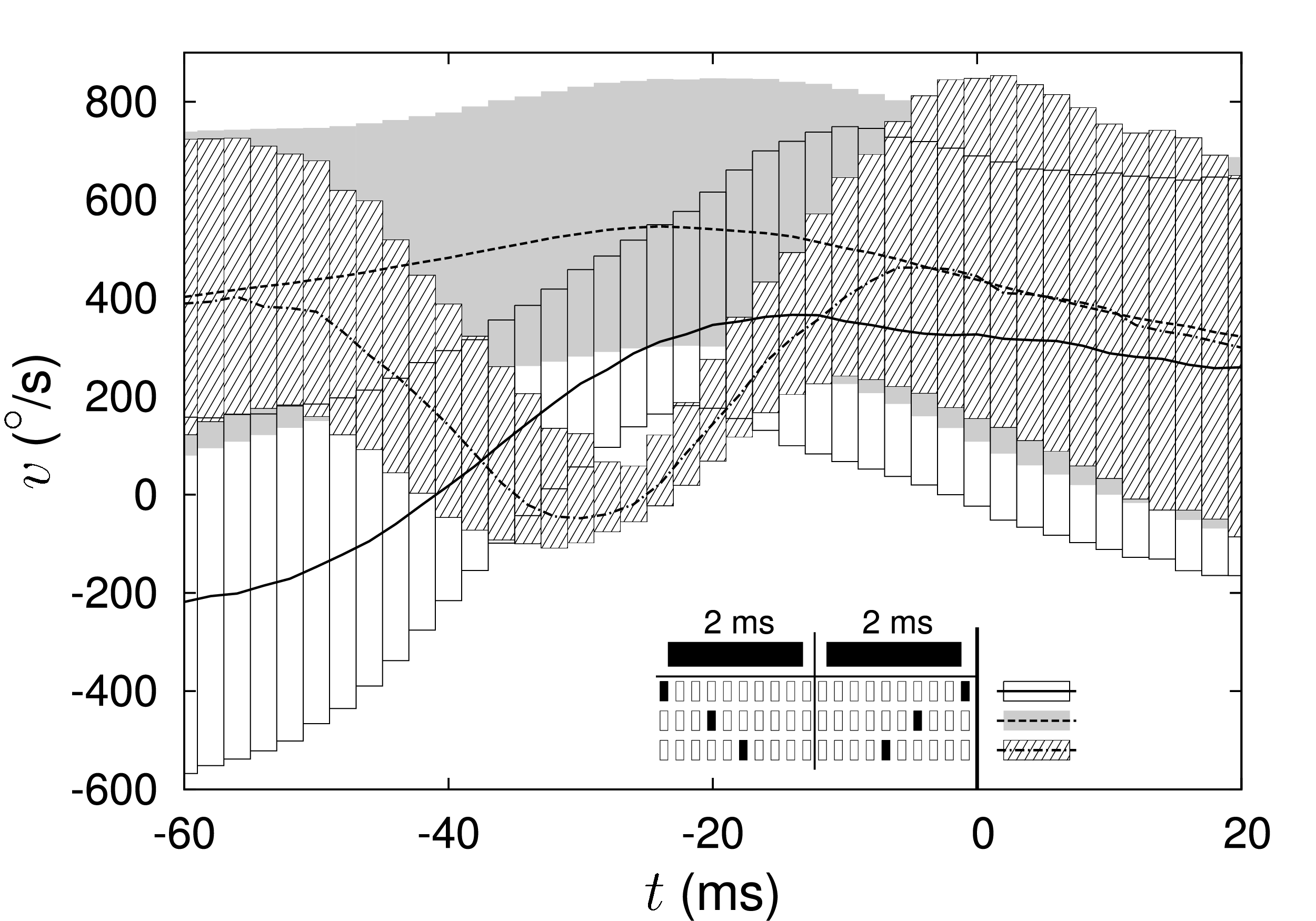}
\hskip 0.25 in
\includegraphics[width = 0.4\linewidth]{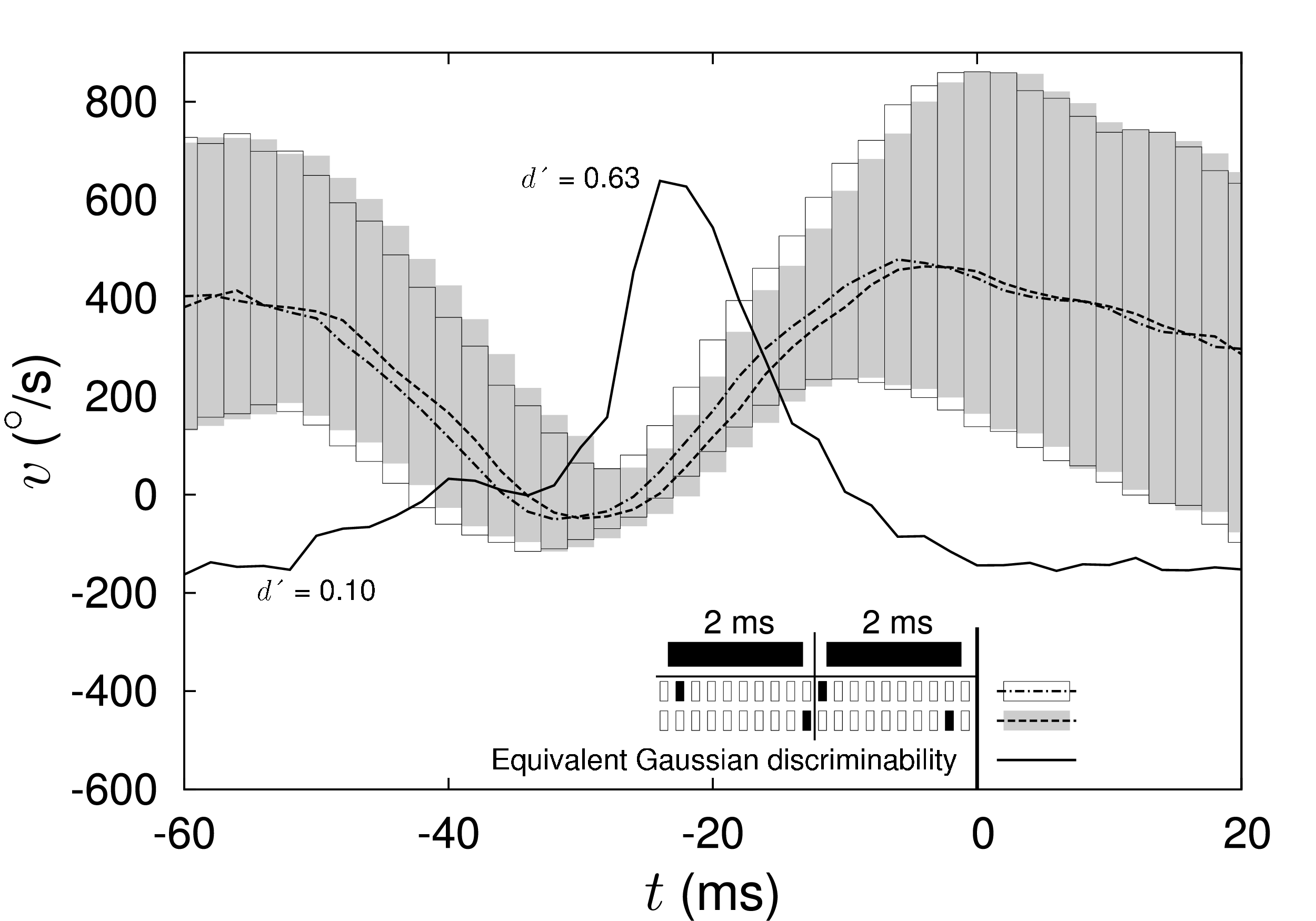}}
\caption{Response conditional ensembles \cite{ruyter+bialek_88}.  We
  consider five different neural responses, all of which are identical
  when viewed at $\tau = 2\,{\rm ms}$ resolution, corresponding to the
  pattern {\tt 11}, spikes in two successive bins.  At left, we
  consider responses which, at higher time resolution, correspond to
  different interspike intervals.  At right, the interspike interval
  is fixed but higher time resolution reveals that the absolute spike
  arrival times differ.  In each case we compute the median motion
  trajectory conditional on the high time resolution response (lines)
  and we indicate the width of the distribution with bars that range
  plus and minus one quartile around the median.  It is clear that
  changes in interspike interval produce changes in the distribution
  of stimulus waveform that are discriminable, since the
  mid--quartiles do not overlap.  Changes in absolute timing are more
  subtle, and so we estimate the conditional distributions of velocity
  at each moment in time using the methods of Ref
  \cite{nemenman+bialek_02}, compute the overlap of these
  distributions, and convert the result into the equivalent
  signal--to--noise ratio $d'$ for discrimination against Gaussian
  noise \cite{green+swets_66}. Note that we compute this
  discriminability using single points in time; $d'$ values based on
  extended segments of the waveforms would be even higher.}
\label{features}
\end{figure*}

The information rate tells us {\sl how much} we can learn about the
sensory inputs by examining the neural response, but it doesn't tell
us {\sl what} we learn.  In particular, we would like to make explicit
the nature of the extra information that emerges as we increase our
time resolution from $\tau = 2\,{\rm ms}$ to $\tau = 0.2\,{\rm ms}$.
To do this, we look at particular ``words'' in a segment of the neural
response, as shown in Fig. \ref{features}, and then examine the motion
trajectories that correspond to these words \cite{ruyter+bialek_88}.
For simplicity, we consider all responses that have two spikes in
successive $2\,{\rm ms}$ bins, that is the pattern {\tt 11} when seen
at $\tau = 2\,{\rm ms}$ resolution.  When we improve our time
resolution to $\tau = 0.2\,{\rm ms}$, some of these responses turn out
to be of the form {\tt 10000000000000000001}, while at the other
extreme some of the responses have the two spikes essentially as close
as is possible given the refractory period, {\tt
  00000100000000100000}.  Remarkably, as we sweep through these subtly
different patterns---which all have the same average spike arrival
time but different interspike intervals---the average velocity
trajectory changes form qualitatively, from a smooth ``on'' (negative
to positive velocity) transition, to a prolonged period of positive
velocity, to a more complex waveform with off and on transitions in
succession.  Examining more closely the distribution of waveforms
conditional on the different responses, we see that these differences
among mean waveforms are in fact discriminable.  Thus, variations in
interspike interval on the millisecond or sub--millisecond scale
represent significantly different stimulus trajectories.

A second axis along which we can ask about the nature of the extra
information at high time resolution concerns the absolute timing of
spikes.  As an example, responses which at $\tau = 2\,{\rm ms}$
resolution are of the form {\tt 11} can be unpacked at $\tau =
0.2\,{\rm ms}$ resolution to give patterns ranging from {\tt
  01000000001000000000} to {\tt 00000000010000000010}, all with the
same interspike interval but with different absolute arrival times.
As shown in Fig \ref{features}, all of these responses code for motion
trajectories with two zero crossings, but the times of these zero
crossings shift as the spike arrival times shift.  Thus, whereas the
times between spikes represent the shape of the waveform, the absolute
arrival time of the spikes mark, with some latency, the time at which
a specific feature of the waveform occurs, in this case a zero
crossing.  Again we find that millisecond and sub--millisecond scale
shifts generate discriminable differences.

The idea that sub--millisecond timing of action potentials could carry
significant information is not new, but the clearest evidence comes
from systems in which the dynamics of the stimulus itself has
significant sub--millisecond structure, as in hearing and
electroreception \cite{carr_93,carr+al_86}. Even for H1, experiments
demonstrating the importance of spike timing at the $\sim 2\,{\rm ms}$
level \cite{strong+al_98,brenner+al_00} could be criticized on the
grounds that the stimuli had unnaturally rapid variations.  It thus is
important to emphasize that, in these experiments, H1 does not achieve
millisecond precision simply because the input has a bandwidth of
kiloHertz; in fact, the stimulus has a correlation time of $\sim
60\,{\rm ms}$ (Fig \ref{redundancy}), and $99.9\%$ of the stimulus
power is contained below $30\,{\rm Hz}$ (Fig \ref{stim}f).

\section{Redundancy reduction}

\begin{figure}
\centerline{\includegraphics[width = 0.9\linewidth]{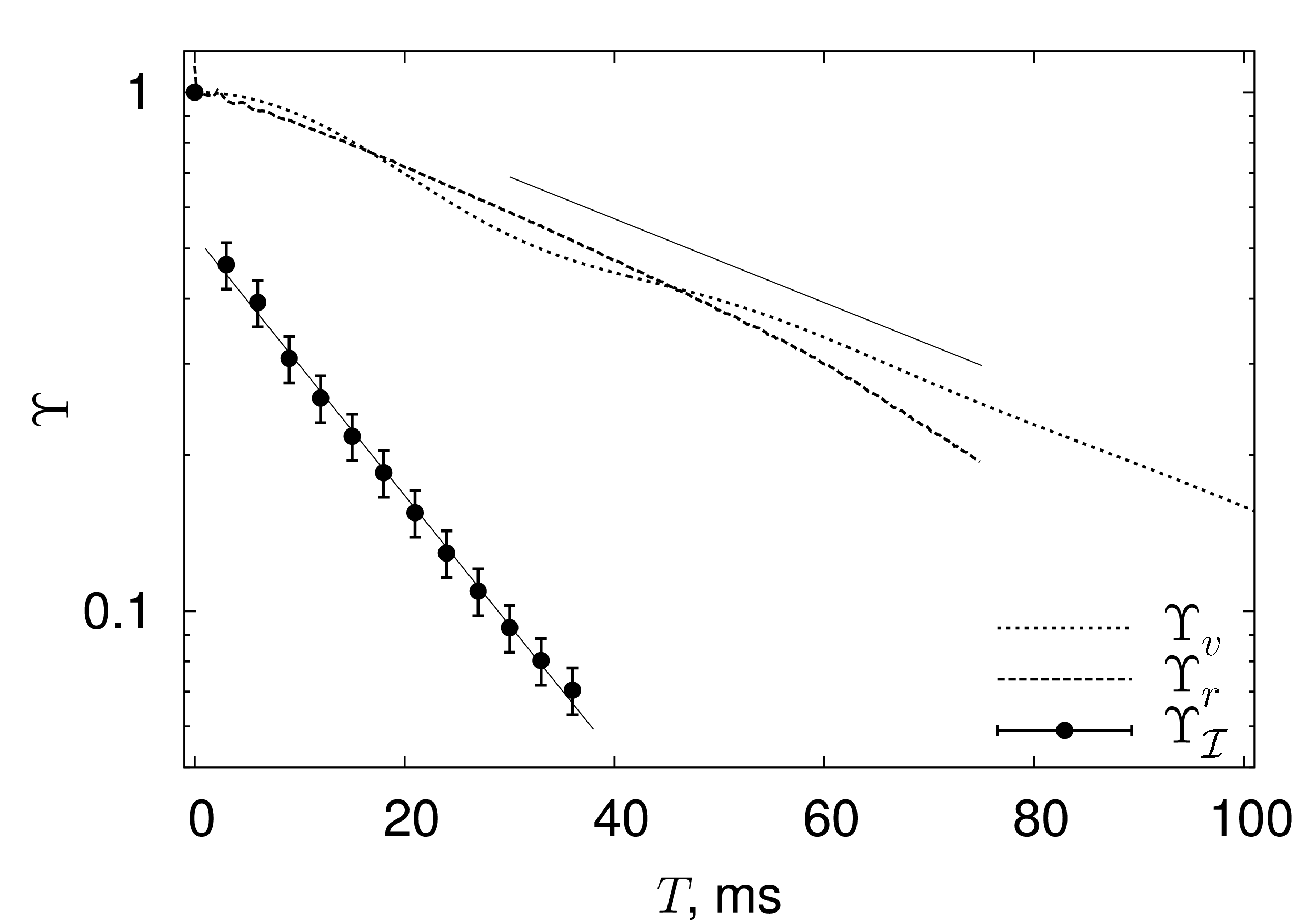}}
\caption{Redundancy reduction in the time domain.  We measure the
  redundancy $\Upsilon_{\cal I} (T,\tau)$ (points with error bars)
  between words of length $T$ in the neural response, as explained in
  the text.  To allow exploration of large $T$ we work at a time
  resolution $\tau = 3\,{\rm ms}$.  The redundancy can be compared to
  correlations in the stimulus $\Upsilon_v = \langle v(t+T)v(t)\rangle
  / \langle v^2\rangle$ (dotted line) or correlations in the spike
  rate $\Upsilon_r = \langle \delta r(t+T)\delta r(t)\rangle /\langle
  (\delta r)^2\rangle$ (dashed line).  Note that the redundancy decays
  rapidly---we show an exponential fit with a time constant of
  $17.3\,{\rm ms}$.  In contrast, the correlations in the stimulus the
  firing rate decay much more slowly---the solid line, for comparison,
  shows an exponential decay with a time constant of $53.4\,{\rm ms}$.
  Correlations in spike rate are calculated from a separate experiment
  on the same cell, with $200$ repetitions of a $10\,{\rm s}$ stimulus
  drawn from the same distribution, that generates more accurate
  estimates of $r(t)$.}
\label{redundancy}
\end{figure}

The long correlation time of these naturalistic stimuli also raises
questions about redundancy---while each spike and interspike interval
can be highly informative, does the long correlation time of the
stimulus inevitably mean that successive spikes carry redundant
information about essentially the same value of the instantaneous
velocity?  Certainly on very short time scales this is true: although
$R_{\rm info}(T,\tau)$ actually increases at small $T$, since larger
segments of the response reveal more informative patterns of several
spikes \cite{brenner+al_00,reinagel+reid_00}, it does decrease at
larger $T$, a sign of redundancy.  On the other hand, the approach to
a constant information rate happens very rapidly: we can measure the
redundancy on time scale $T$ by computing $\Upsilon_I(T,\tau ) =
2I(T,\tau )/I(2T,\tau ) - 1$, where $\Upsilon = 0$ means that
successive windows of size $T$ provide completely independent
information, and $\Upsilon = 1$ means that they are completely
redundant.  As shown in Fig \ref{redundancy}, $\Upsilon_I(T,\tau )$
decays rapidly, on a time scale of less than $20\,{\rm ms}$.  In
contrast, correlations in the stimulus decay much more slowly, on the
$\sim 60\,{\rm ms}$ time scale.  Further, we can compute at each
moment of time the spike rate $r(t)$, and this has a correlation time
comparable to the stimulus itself, suggesting that the decorrelation
of information is more subtle than a simple filtering of the stimulus.

\section{Bit rates and photon counting rates}

The ability of the fly's visual system to mark features of the
stimulus with millisecond precision, even when the stimulus
correlation time is $\sim 60 \,{\rm ms}$, depends on having access to
a representation of visual motion with very high signal--to--noise
ratio.  Previous work has suggested that this system can estimate
motion with a precision close to the limits set by noise in the
photoreceptors \cite{bialek+al_91,ruyter+bialek_95}, which is
dominated by photon shot noise
\cite{ruyter+laughlin_96a,ruyter+laughlin_96b}.  The present
experiments, however, are done under very different conditions:
velocities of motion are much larger, the fly's eye is stimulated over
a much larger area, and light intensities outdoors are much larger
than generated by laboratory displays.  During the course of our
experiments we monitor the light intensity at zenith, using a detector
matched to the properties of the fly photoreceptors (see Methods);
from these measurements we estimate that the mean light intensity
corresponds to $1.56\times 10^6\,{\rm photon/s}$ per photoreceptor,
which is near the limit of the photoreceptor's dynamic range for
photon counting.  Is it possible that photon counting statistics still
are relevant even at these high rates?

Because the experiments are done outdoors, there are small
fluctuations in light intensity from trial to trial as clouds drift by
and obscure the sun.  Although the dynamic range of these fluctuations
is less than a factor two, the arrival times of individual spikes
(e.g., the ``first spike'' after $t=1.75\,{\rm s}$ in Fig
\ref{big_expt}) have correlation coefficients of up to $\rho = -0.42$
with the light intensity, with the negative sign indicating that
higher light intensities lead to earlier spikes.  One might see this
effect as a failure of the system to adapt to the overall light
intensity, but it also suggests that some of what we have called noise
really represents a response to trial--by--trial variations in
stimulus conditions. Indeed, a correlation between light intensity and
spike time means that the noise entropy $S_n(T,\tau|t)$ in windows
which contain these spikes is smaller than we have estimated because
some of the variability can be ascribed to stimulus variation.

More subtly, if photon shot noise is relevant, we expect that on
trials with higher light intensity the neuron will actually convey
more information about the trajectory of motion.  We emphasize that
this is a delicate question.  To begin, the differences in light
intensity are small, and we expect (at most) proportionately small
effects.  Further, as the light intensity increases, the total spike
rate increases, and this increases both the total entropy and the
noise entropy.  To ask if the system uses the more reliable signal at
higher light intensities to convey more information we have to
determine which of these increases is larger.

To test the effects of light intensity on information transmission
(see Methods for details), we divide the trials into halves based on
the average light intensity over the trial, and we try to estimate the
information rates in both halves; the two groups of trials differ by
just $3\%$ in their median light intensities.  Since cutting the
number of trials in half makes our sampling problems much worse, we
focus on short segments of the response ($T=6\,{\rm ms}$) at high time
resolution ($\tau = 0.2\,{\rm ms}$); note that these are still
``words'' with 30 letters.  For this case we find that for the trials
with higher light intensities the information about the motion
stimulus is larger by $\Delta = 0.0204 \pm 0.0108 \,{\rm bits}$, which
is small but significant at the $94\%$ confidence level.  We find
differences with the same sign for all accessible combinations of $T$
and $\tau$, and the overall significance of the difference thus is
much larger.  Note that since we are analyzing $T=6\,{\rm ms}$
windows, this difference corresponds to $\Delta R \sim 3\,{\rm
  bits/s}$, $1-2\%$ of the total (cf Fig \ref{ent+inf_results}). Thus
even at rates of more than one million photons per second per receptor
cell, small increases in photon flux produce significant changes in
the transmission of information about the visual input.

\section{Conclusion}

To summarize, we have found that under natural stimulus conditions the
fly visual system generates spikes and interspike intervals with
extraordinary temporal precision.  As a consequence, the neural
response carries a substantial amount of information that is available
only at sub--millisecond time resolution.  At this high resolution,
absolute spike timing is informative about the time at which
particular stimulus features occur, while different interspike
intervals provide a rich representation of distinguishable stimulus
features.  These results provide a clear demonstration that the visual
system uses sub--millisecond timing to provide a richer representation
of the natural sensory world, at least in this corner of the fly's
brain.  In addition, the data provide support for the idea that the
system performs efficiently both in the tasks of estimation and
coding, making use of the extra signal--to--noise provided by
increased photon flux and reducing the redundancy of the stimulus as
it is transformed into spikes.  Finally, we note that our ability to
reach these conclusions depends not just on new experimental methods
that allow us to generate truly naturalistic stimuli
\cite{lewen+al_01}, but critically on new mathematical methods that
allow us to analyze neural responses quantitatively even when it is
impossible for us to sample the distribution of responses exhaustively
\cite{nemenman+al_02,nemenman+al_04}.  We expect that these sorts of
mathematical tools will become even more critical for neuroscience in
the future.

\acknowledgments{We thank J Miller, KD Miller and the participants in
  the NIPS 2003 Workshop on Entropy estimation for helpful
  discussions.  This work was supported in part by grants from the
  National Science Foundation (PHY99--07949, ECS--0425850,
  IIS--0423039), the Department of Energy under contract
  DE--AC52--06NA25396, and the Swartz Foundation.  Early stages of
  this work were done when all the authors were at the NEC Research
  Institute.  IN thanks the Kavli Institute for Theoretical Physics
  and Columbia University for their support during this work, and WB
  thanks the Center for Theoretical Neuroscience at Columbia
  University for its hospitality.}

\appendix

\section{Methods}

{\bf Neural recording and stimulus generation.}  H1 was recorded
extracellularly by a short ($12\,{\rm mm}$ shank length) tungsten
electrode (FHC). The signal was preamplified by a differential
bandpass amplifier based on the INA111. After amplification by a
second stage samples were digitized at $10\,{\rm kHz}$ by an AD
converter (National Instruments DAQCard--AI--16E--4, mounted in a
Fieldworks FW5066P ruggedized laptop). In off line analysis, the
analog signal was digitally filtered by a template derived from the
average spike waveform. Spikes were then time stamped by interpolating
threshold crossing times. The ultimate precision of this procedure is
limited by the signal to noise ratio in the recording; for typical
conditions this error is estimated to be $50-100\,\mu{\rm s}$.  Note
that we analyze spike trains down to a precision of $\tau =
200\,\mu{\rm s}$, so that some saturation of information at this high
time resolution may actually result from instrumental limitations.
The experiments were performed outside in a wooded environment, with
the fly mounted on a stepper motor with vertical axis. The speed of
the stepper motor was under computer control, and could be set at
$2\,{\rm ms}$ intervals.  The DAQ card generates a clock signal at
$500\,{\rm Hz}$ in synchrony with the master clock which calibrates
the neural recording.  As explained in the legend to Fig
\ref{big_expt}, each tick of the clock drives the stepper motor
through an amount determined by reading the stimulus file stored on a
dedicated computer.  The motor (SIG--Positec RDM566/50 stepper motor,
104 pulses per revolution) is driven by a controller (SIG--Positec
Divistep D331.1), which in turn receives pulses at a frequency divided
down from a free running $8\,{\rm MHz}$ clock; the stimulus velocity
is represented by the divisor for the pulse frequency.  In this way,
the stepper motor is driven in each $2\,{\rm ms}$ period, in strict
synchrony with the data acquisition clock, by steps that are evenly
spaced.  This design was chosen to minimize the effects of discrete
steps and to maximize the reliability of all timing measurements. To
stabilize temperature the setup was enclosed by a transparent
plexiglass cylinder (radius $15\,{\rm cm}$, height $28\,{\rm cm}$),
with a transparent plexiglass lid.

{\bf Monitoring light intensity and controlling temperature.} The air
temperature in the experimental enclosure was regulated by a Peltier
element fitted with heat vanes and fans on both sides for efficient
heat dispersal, driven by a custom built feedback controller. The
temperature could be set over a range from approximately five degrees
below to fifteen degrees above ambient temperature, and the controller
stabilized temperature over this range to within about a degree. In
the experiments described here, temperature was $23\pm1^\circ\,{\rm
  C}$.  An overall measure of light intensity was obtained by
monitoring the current of a photodiode (Hamamatsu) enclosed in a
diffusing ping pong ball. The photodiode signal was amplified by a
logarithmic amplifier operating over five decades. The photodiode was
located $\sim 50\,{\rm cm}$ from the fly, and in the experiments the
setup was always placed in the shade. The photodiode measurement was
intended primarily to get a rough impression of relative light
intensity fluctuations. However, to relate these measurements to
outside light levels, before the start of each experiment a separate
calibration measurement of zenith radiance was taken using a
calibrated light intensity meter. To relate this measurement to fly
physiology, the radiance reading was converted to an estimated
effective fly photoreceptor photon rate. The reading of the photodiode
was roughly proportional to the zenith intensity reading, with a
proportionality factor determined by the placement of the setup and
the time of day. To obtain a practical rule of thumb, the photodiode
readings were converted to equivalent zenith photon flux values, using
the current to zenith intensity conversion factor established at the
beginning of the experiment. During the experiments the photodiode
current was sampled at $1\,{\rm s}$ intervals.

{\bf Repeated stimuli.}  In their now classical experiments, Land and
Collett measured the trajectories of flies in free flight
\cite{land+collett_74}; in particular they reported the angular
position (orientation) of the fly vs time, from which we can compute
the angular velocity $v(t)$.  The short segments of individual
trajectories shown in the published data have a net drift in angle, so
we include both the measured $v(t)$ and $-v(t)$ as parts of the
stimulus.  We use the trajectories for the two different flies in Fig
4 of Ref \cite{land+collett_74}, and graft all four segments together,
with some zero padding to avoid dramatic jumps in velocity, generating
a stimulus that is 5 seconds in duration and has zero drift so that
repetition of the angular velocity vs time also repeats the angular
position vs time.  Since Land and Collett report data every $20\,{\rm
  ms}$, we interpolate to generate a signal that drives the stepper
motor at $2\,{\rm ms}$ resolution; interpolation is done using the
MATLAB routine {\tt interp}, which preserves the bandlimited nature of
the original signal and hence does not distort the power spectrum.

{\bf Nonrepeated stimulus.}  To analyze the full entropy of neural
responses, it is useful to have a stimulus that is not repeated.  We
would like such a stimulus to match the statistical properties of
natural stimulus segments described above.  To do this, we estimate
the probability distribution $P[v(t+\Delta t)| v(t)]$ from the
published trajectories, where $\Delta t = 20\,{\rm ms}$ is the time
resolution, and then use this as the transition matrix of a Markov
process from which we can generate arbitrarily long samples; our
nonrepeated experiment is based on a $990\,{\rm s}$ trajectory drawn
in this way.  The resulting velocity trajectories will, in particular,
have exactly the same distributions of velocity and acceleration as in
the observed free flight trajectories.  Although the real trajectories
are not exactly Markovian, our Markovian approximation also captures
other features of the natural signals, for example generating a
similar number of velocity reversals per second.  Again we interpolate
these trajectories to obtain a stimulus at $2\,{\rm ms}$ resolution.

{\bf Entropy estimation in a model problem.}  The problem in Fig
\ref{errors} is that of a potentially biased coin.  Heads appear with
probability $p$, and the probability of observing $n$ heads out of $N$
flips is
\begin{equation}
P_N (n|p) \propto p^n (1-p)^{N-n} .
\end{equation}
If we observe $n$ and try to infer $p$, we use Bayes' rule to construct
\begin{equation}
P_N (p|n) = P_N(n|p) {{{\cal P}(p)}\over{P_N(n)}}\propto {\cal P}(p) p^n (1-p)^{N-n} ,
\end{equation}
where ${\cal P}(p)$ is our prior and $P_N(n) = \int_0^1 dp\,
P_N(n|p){\cal P}(p)$.  Given this posterior distribution of $p$ we can
calculate the distribution of the entropy,
\begin{equation}
S(p) = -p\log_2 (p) - (1-p)\log_2 (1-p) .
\end{equation}
We proceed as usual to define a function $g(S)$ that is the inverse of
$S(p)$, that is $g(S(p)) = p$; since $p$ and $1-p$ give the same value
of $S$, we choose $0< g \leq 0.5$ and let $\tilde g (S) = 1-g(S)$.
Then we have
\begin{equation}
P_N (S|n) = \left[ P_N (p=g(S)|n) + P_N(p = \tilde g(S)|n)\right]
{\bigg |} {{dg(S)}\over{dS}}{\bigg |} .
\end{equation}
From this distribution, we can estimate a mean $\bar S_N (n)$ and a
variance $\sigma^2 (n,N)$ in the usual way.  What interests us is the
difference between $\bar S_N (n)$ and the true entropy $S(p)$
associated with the actual value of $p$ characterizing the coin; it
makes sense to measure this difference in units of the standard
deviation $\delta S(n,N)$.  Thus we compute
\begin{equation}
\langle (S' - S_0)/\delta S\rangle \equiv 
\sum_{n=0}^N P_N(n|p) \left[ {{\bar S_N(n) - S(p)}\over{\delta S(n,N)}}\right] ,
\end{equation}
and this is what is shown in Fig \ref{errors}.  We consider two cases.
First, a flat prior on $p$ itself, so that ${\cal P}(p) = 1$.  Second,
a flat prior on the entropy, which corresponds to
\begin{eqnarray}
{\cal P}(p) &=& {1\over 2}{\bigg |} {{dS(p)}\over{dp}}{\bigg |}\\
&=& {1\over 2}{\bigg |} \log_2\left( {{1-p}\over p}\right) {\bigg |} .
\end{eqnarray}
Note that this prior is (gently) diverging near the limits $p=0$ and
$p=1$, but all the expectation values that we are interested in are
finite.

{\bf Entropy estimation: General features.}  Our discussion here
follows Refs \cite{nemenman+al_02,nemenman+al_04} very closely.
Consider a set of possible neural responses labeled by ${\rm i} = 1,
2, \cdots , K$.  The probability distribution of these responses,
which we don't know, is given by ${\bf p}\equiv \{p_{\rm i}\}$.  A
well studied family of priors on this distribution is the Dirichlet
prior, parameterized by $\beta$,
\begin{equation}
{\cal P}_\beta ({\bf p}) = {1\over{Z(\beta ; K)}}\left[ \prod_{{\rm i}=1}^K p_{\rm i}^{\beta -1}\right] \delta\left(\sum_{{\rm i} =1}^K p_{\rm i} -1\right) .
\end{equation}
Maximum likelihood estimation, which identifies probabilities with
frequencies of occurrence, is obtained in the limit $\beta\rightarrow
0$, while $\beta \rightarrow 1$ is the natural ``uniform'' prior.
When $K$ becomes large, almost any $\bf p$ chosen out of this
distribution has an entropy $S=-\sum_{\rm i} p_{\rm i}\log_2 p_{\rm
  i}$ very close to the mean value,
\begin{equation} 
\bar S (\beta ; K) = \psi_0 (K\beta + 1) - \psi_0 (\beta + 1),
\end{equation}
where $\psi_0 (x) = d\log_2\Gamma(x)/dx$, and $\Gamma(x)$ is the gamma
function.  We therefore construct a prior which is approximately flat
on the entropy itself by a continuous superposition of Dirichlet
priors,
\begin{equation}
{\cal P}({\bf p}) = \int d\beta {{\partial\bar S (\beta; K)}\over {\partial\beta}} {\cal P}_\beta ({\bf p}) ,
\end{equation}
and we then use this prior to perform standard Bayesian inference.  In
particular, if we observe each alternative $\rm i$ to occur $n_{\rm
  i}$ times in our experiment, then
\begin{equation}
P(\{n_{\rm i}\} | {\bf p}) \propto \prod_{{\rm i}=1}^K p_{\rm i}^{n_{\rm i}} ,
\label{Pngivenp}
\end{equation}
and hence by Bayes' rule
\begin{equation}
P({\bf p}| \{n_{\rm i}\}  ) \propto \left[ \prod_{{\rm i}=1}^K p_{\rm i}^{n_{\rm i}}\right] {\cal P}({\bf p}) .
\end{equation}
Once we normalize this distribution we can integrate over all $\bf p$
to give the mean and the variance of the entropy given our data
$\{n_{\rm i}\}$.  In fact, all the integrals can be done analytically
except for the integral over $\beta$.  Software implementation of this
approach is available from {\tt http://nsb-entropy.sourceforge.net/}.
This basic strategy can be supplemented in cases where we have prior
knowledge about the entropies.  In particular, when we are trying to
estimate entropy in ``words'' of increasing duration $T$, we know that
$S(T,\tau ) \leq S(T',\tau) + S(T-T',\tau)$ for any $T'$, and thus it
makes sense to constrain the priors at $T$ using the results from
smaller windows, although this is not critical to our results.  We
obtain results at all integer values of $T/\tau$ for which our
estimation procedure is stable (see below) and use cubic splines to
interpolate to non--integer values as needed.

{\bf Entropy estimation: Details for total entropy.}  There are two
critical challenges to estimating the entropy of neural responses to
natural signals.  First, the overall distribution of (long) words has
a Zipf--like structure (Fig \ref{ent+inf_results}b), which is
troublesome for most estimation strategies and leads to biases
dependent on sample size.  Second, the long correlation times in the
stimulus mean that, successive words `spoken' by the neuron are
strongly correlated, and hence it is impossible to guarantee that we
have independent samples, as assumed implicitly in Eq
(\ref{Pngivenp}).  We can tame the long tails in the probability
distribution by partitioning the space of responses, estimating
entropies within each partition, and then using the additivity of the
entropy to estimate the total.  We investigate a variety of different
partitions, including (a) no spikes vs.\ all other words, (b) no
spikes, all words with one spike, all words with two spikes, etc., (c)
no spikes, all words with frequencies of over 1000, and all other
words.  Further, for each partitioning, we follow \cite{strong+al_98}
and evaluate $S(T,\tau)$ for data sets of different sizes $\alpha N$,
$0<\alpha\le 1$. Note that by choosing fractions of the data in
different ways we can separate the problems of correlation and sample
size.  Thus, to check that our estimates are stable as a function of
sample size, we choose contiguous segments of experiment, while to
check for the impact of correlations we can `dilute' our sampling so
that there are longer and longer intervals between words.  Obviously
there are limits to this exploration (one cannot access large, very
dilute samples), but as far as we could explore the impact of
correlations on our estimates is negligible once the samples sizes are
sufficiently large.  For the effects of sample size we look for
behavior of the form $S(\alpha)=S_\infty +S_1/\alpha+S_2/\alpha^2$ and
take $S_\infty$ as our estimate of $S(T,\tau)$, as in Ref
\cite{strong+al_98}.  For all partitions in which the the most common
word (silence) is separated from the rest, these extrapolated
estimates agree and indicate negligible biases at all combinations of
$\tau$ and $T$ for which the $1/\alpha^2$ term is negligible compared
to the $1/\alpha$ (that is, $\tau\ge0.5$ ms at $T\le 25$ ms). For
smaller $\tau$, estimation fails at progressively smaller $T$, and to
obtain an entropy rate for large $T$ we extrapolate to $\tau/T\to0$
using
\begin{equation}
{1\over T} S(T,\tau )  = {\cal S} (\tau) + A (\tau/T) + B (\tau /T)^2  ,
\end{equation}
where ${\cal S}(\tau )$ is our best estimate of the entropy rate at
resolution $\tau$.  All fits were of high quality, and the resulting
error bars on the total entropy are negligible compared to those for
the noise entropy.  In principle, we could be missing features of the
code which appear {\em only} when we use high resolution for very long
words, but this unlikely scenario is almost impossible to exclude by
any means.

{\bf Entropy estimation: Details for noise entropy.}  Putting error
bars on the noise entropy averaged over time is more difficult because
these should include a contribution from the fact that our finite
sample over time is only an approximation to the true average over the
underlying distribution of stimuli.  Most seriously, the entropies are
very different in epochs that have net positive or negative
velocities.  Because of the way that we constructed the repeated
stimulus, $v(t) = -v(t+T_0)$, with $T_0 = 2.5\,{\rm s}$; thus if we
compute $S_n(T,\tau|t)+ S_n(T,\tau|t+T_1)$ with $T_1\approx T_0$, this
fluctuates much less as a function of $t$ than the entropy in an
individual slice.  Because our stimulus has zero mean, every slice has
a partner under this shift, and the small difference between $T_0$ and
$T_1$ takes account of the difference in latency between responses to
positive and negative inputs.  A plot of $S_n(T,\tau|t)+
S_n(T,\tau|t+T_1)$ vs time $t$ has clear dips at times corresponding
to zero crossings of the stimulus, and we partition the data at these
points.  We derive error bars on the mean noise entropy $\langle
S_n(T,\tau|t)\rangle_t$ by a bootstrap--like method, in which we
construct samples by randomly sampling with replacements from among
these blocks, jittering the individual entropies $S_n(T,\tau|t)$ by
the errors that emerge from the Bayesian analysis of individual
slices.  As with the total entropy we extrapolate to otherwise
inaccessible combinations of $T$ and $\tau$, now writing
\begin{eqnarray}
{1\over T} \langle S_n(T,\tau|t)\rangle_t
&=& {\cal S}_n (\tau) + A (\tau/T) + B (\tau /T)^2 \nonumber\\
&&\,\,\,\,\,\,\,\,\,\,+ C\cos(2\pi T/\tau_0) 
\end{eqnarray}
and fitting by weighted regression.  Note that results at different
$T$ but the same value of $\tau$ are strongly correlated, and so the
computation of $\chi^2$ is done using the full (non--diagonal)
covariance matrix.  The periodic term is important at small $\tau$,
where we can see structure as the window size $T$ crosses integer
multiples of the average interspike interval, $\tau_0 = 2.53\,{\rm
  ms}$.  Error estimates emerge from the regression in the standard
way, and all fits had $\chi^2 \sim 1$ per degree of freedom.

{\bf Impact of photon flux on information rates.}  Since there are no
responses to repeated and unrepeated stimuli recorded at exactly the
same illuminations, we use the data from the repeated experiment to
evaluate both the noise entropy and the total entropy.  We expect that
we are looking for small differences, so we tighten our analysis by
discarding the first two trials, which are significantly different
from all the rest (presumably because adaptation is not complete), as
well as excluding the epochs in which the stimulus was padded with
zeroes.  The remaining 98 trials are split into two groups of 49
trials each with the highest and the lowest ambient light levels.  We
can then estimate the total entropy $S^{(h,l)}(T,\tau )$ for the high
$(h)$ and low $(l)$ intensity groups of trials, and similarly for the
noise entropy in each slice at time $t$, $S_n^{(h,l)}(T,\tau |t)$.  As
noted above, assigning error bars is clearer once we form quantities
that are balanced across positive and negative velocities, and we do
this directly for the difference in noise entropies,
\begin{eqnarray}
\Delta S_n (T,\tau; t) &=& [S_n^{(h)}(T,\tau |t) + S_n^{(h)}(T,\tau |t+T_1)]\nonumber\\
&&\,\,\,\,\,\,\,\,\,\,
- [S_n^{(l)}(T,\tau |t) + S_n^{(l)}(T,\tau |t+T_1')],\nonumber\\
&&
\end{eqnarray}
where we allow for a small difference in latencies ($T_1 - T_1'$)
between the groups of trials at different intensities.  We find that
$\Delta S_n (T,\tau; t)$ has a unimodal distribution and a correlation
time of $\sim1.4$ ms, which allows for an easy evaluation of the
estimation error.

\end{document}